# Realistic Multimedia Tools based on Physical Models: I. The Spectrum Analyzer and Animator (SA$^2$)

Ioannis Pachoulakis

Department of Applied Informatics & Multimedia, TEI of Crete
and
Centre for Technological Research of Crete
ip@epp.teicrete.gr

## Abstract

The present sequence of two articles reports on a custom-built toolkit implementing a technique similar to multi-directional medical tomography to simulate and visualize the composite 3D structure of winds from hot close double stars. In such hot binaries, the light sources scanning and probing the composite wind volume are the bright "surfaces" (photospheres) of the individual stars. Then, as the Keplerian orbit is traced out and the geometry presented to the observer varies, each star constitutes an analyzer upon its companion's wind. In contrast to medical tomography, however, these targets are too far to be resolved spatially so we resort to modeling the ultraviolet (UV) spectral lines of certain wind ions (e.g., N+4, Si+3, C+3) whose shapes vary with Keplerian phase as the stars revolve around their common centre of mass. The flagships of the toolkit are the Spectrum Analyzer and Animator (SA$^2$) and the Binary 3D Renderer (B3dR). The SA$^2$ is the subject of the present article (paper I). It automates (a) the derivation of light curves from the observed spectra and (b) the generation of synthetic binary wind-line profiles which reproduce the morphologies and variabilities of the observed wind profiles. The second tool, the B3dR is discussed in paper II.

## 1. Introduction

Most stars in the night sky are not single stars but are, in fact, composed of two stars which revolve around their common centre of mass. In spectroscopic binaries the stars are so close to each other that they remain unresolved even through a telescope: their binary nature is recovered only from their spectra thanks to the Doppler Effect. Of these spectroscopic binaries, the subset of eclipsing binaries is of the greatest importance, as their study discloses accurate masses and radii, which provide the prime test bed of stellar models. Hot, massive (O and B) Main Sequence stars are at least 10 times more massive and much hotter (T ≥ 30,000 K) than the sun, so that material from their surface is blown away by radiation pressure, forming stellar winds composed of charged particles (protons, electrons and ions) accelerated to supersonic speeds.

Now, as the Keplerian orbit is traced out, the geometry presented to the observer varies and each star constitutes an analyzer upon its companion's wind, probing that wind's structure in a process reminiscent of tomography. Indeed, stellar wind studies use Doppler-resolved spectral signatures to reconstruct the composite wind from a sequence of spectra recorded as the component stars coast along their orbits. The modeling process detailed in this article uses the spectral signatures of prominent wind ions such as those of $N^{+4}$, $Si^{+3}$, $C^{+3}$ which trace the wind structure forming the well-known wind lines of NV, SiIV and CIV respectively, and whose profiles generally vary as the stars revolve and as wind inhomogeneities form, propagate and evolve.

Using UV spectra secured by the International Ultraviolet Explorer (IUE) satellite at various orbital phases, this article shows how to model these spectral lines, reconstruct the 3D composite wind structure and simulate the motion of the system around the common center of mass. The output is realistic content that can be incorporated in multimedia applications to increase the effectiveness and communication impact of research results. The functional keyword "realistic" means content that is based on physical modeling and, therefore, is physically sound.

## 2. Synthetic Binary Wind-Line Profiles

The showcase of our analysis is HD159176, which is the brightest object (V = + 5.68, B-V = +0.04) in the Milky Way open cluster NGC 6383 (Lloyd-Evans, 1978), and is classified as O7V + O7V by Hiltner, Garrison and Schild (1969) and as O6V + O6V by Levato (1975). The observational material for HD159176, the derived orbital elements as well as some first results in visualizing its geometry have been presented and extensively discussed elsewhere (Pachoulakis, 2006a,b). In the present section, we show how to model the wind spectral lines of HD159176.

The SEI (Sobolev with Exact Integration) method for calculating line profiles of spherically symmetric winds from single stars is described in Lamers, Cerruti-Sola and Perinoto (1987), hereafter LCP, and has been applied to O star winds (e.g., Groenewegen and Lamers, 1989, Groenewegen, Lamers and Pauldrach 1989) and to outflows from planetary nebulae (e.g., Perinotto, Cerutti-Sola and Lamers, 1989). The method allows for radiative interaction between the blue and red members of a doublet such as NV, SiIV and CIV and was applied in this article to generate synthetic binary wind-line model profiles as follows.

For a two-level atom/ion, the source function in SEI is written as:

$$S_\nu = \frac{\delta_c(r) \cdot I_\nu^* + \varepsilon \cdot B_\nu(r)}{\delta + \varepsilon},$$

where $I_\nu^*$ is the intensity at frequency ν of the radiation leaving the stellar photosphere, $B_\nu$ is the Plank function in the wind, δ is the escape probability for line photons, $\delta_c$ is the penetration probability for continuum photons, and ε is the ratio of collisional to radiative de-excitations. The collisional term ε for the source function is negligible for resonance lines in O and B stars and is set to zero, accordingly. The relevant parameters employed by SEI to generate a model profile appropriate for a stellar wind from an O or B star are discussed directly below.

For a single star, the wind speed for a given ion is modeled by a β-type law:

$$w(r) = w_0 + (1 - w_0) \cdot (1 - 1/r)^\beta,$$

where w(r) is the wind speed v(r) normalized to $v_\infty$, i.e., $w(r) = v(r) / v_\infty$, and $w_0$ is the normalized wind speed at the wind base. The parameter $w_{Gauss} \equiv v_{Gauss} / v_\infty$ mimics the occurrence of small-scale perturbations of the velocity law by introducing, at distance r, a Gaussian distribution of speeds centered on w(r) with a half-width $w_{Gauss}$. $A_{photB}$ and $A_{photR}$ are the photospheric depths (in normalized flux) of the blue and red components of the doublet, respectively. Similarly, $W_{photB}$ and $W_{photR}$ are the photospheric widths of the blue and red members of the doublet. Because SEI does not solve the statistical equilibrium equations in the wind, an optical depth law is specified which has the following form (Equation 40 in LCP):

$$\frac{d\tau}{dw} = \frac{T_{Tot}}{I}\left(\frac{w}{w_1}\right)^{\alpha_1}\left[1-\left(\frac{w}{w_1}\right)^{1/\beta}\right]^{\alpha_2}$$

The optical depth law so specified is quite versatile, with $\alpha_{1,2}$ acting as shaping parameters for the functional dependence of $d\tau/dw$ on wind speed w and, via the velocity law, for the dependence of $d\tau/dr$ on distance r. Furthermore, as observable line formation does not necessarily extend to $v_\infty$, a parameter $w_1$ is introduced, which defines the extent of the wind in velocity space, in the sense that line formation ceases beyond $w = w_1$. $T_{photB}$ and $T_{photR}$ are the integrated radial optical depths for the blue and red components of the doublet, respectively. Finally, the integral

$$I = \int_{w_0/w_1}^{1} y^{\alpha_1}(1-y^{1/\beta})^{\alpha_2} dy$$

is introduced for normalization, so that

$$\int_{w_0}^{w_1} \frac{d\tau}{dw} dw = T_{Tot}$$

Figure 1 displays the build-up of wind speed (left panel) and the drop-off of optical depth (right panel) from the surface of a single star with the characteristics of the primary star of HD159176 using parameter values from Pachoulakis (1996).

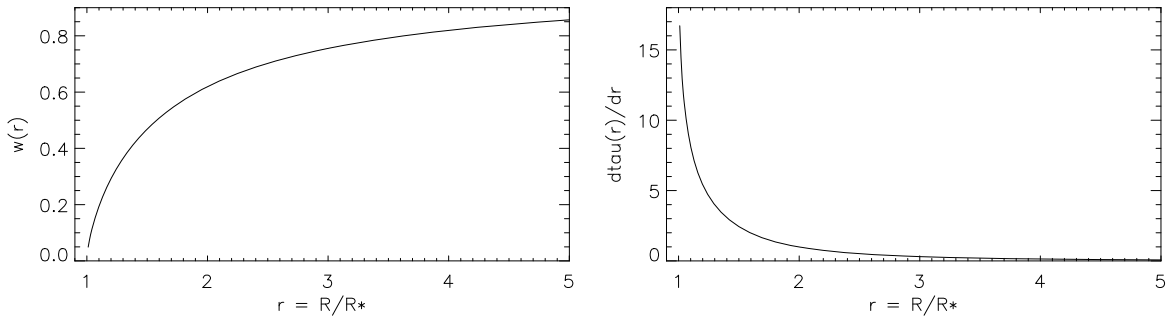

**Figure 1: The build-up of wind speed (left panel) and the drop-off of optical depth (right panel).**

In SEI, $S_v$ is solved in the Sobolev approximation but the equation of transfer is integrated exactly along lines of sight with impact parameters $p \geq 0$ using equations (28) and (29) in LCP. The left panel of Figure 2 shows a schematic of an extended wind around a single star and the right panel shows the SEI single-star CIV wind-line profile (continuous curve). For any line of sight to a distant observer, the impact parameter p is the normalized distance from the star center. Then, as seen from Earth, the SEI model profile is the sum of:

- The pure emission contribution (dashed line in right panel) from the part of the wind with p ≥ 1 (horizontally hatched projected area in left panel), plus
- The pure absorption contribution (dotted line in right panel) from the part of the wind directly in front of the star with p < 1 (cross-hatched projected area in left panel) as material accelerates in the direction to the observer. (The reader may note that the contribution to line formation of the part of the wind with p < 1 which is behind the star is zero as it is occulted by the stellar disk.)

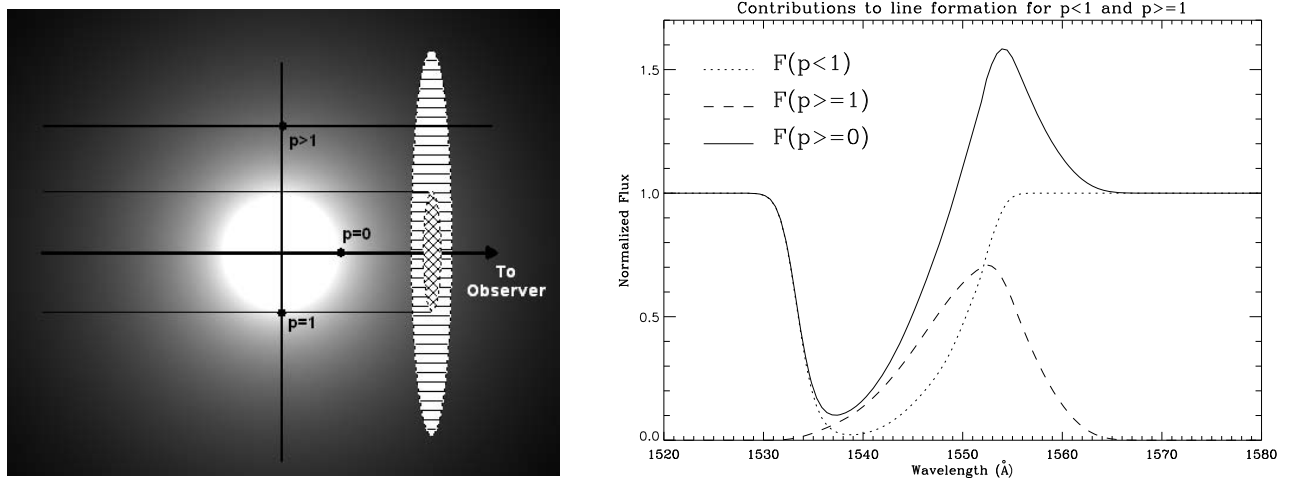

**Figure 2: Schematic of an extended wind around a single star (left panel) and SEI wind-line profile (continuous curve in the right panel).**

For the extended, optically thick winds from early O-type stars like HD159176 both contributions to line formation (i.e., for p ≥ 1 and for p < 1) are significant, the net result being the characteristic P-Cygni profile shown in the right panel of Figure 2 which combines absorption and emission. On the other hand, for late-O or early-B stars, most of the wind-line formation takes place within a shell of thickness just a small fraction of the stellar radius, so that the contribution to line formation for p ≥ 1 dominates that for p < 1 and the net profile is an absorption feature.

## 3. THE SPECTRUM ANALYZER AND ANIMATOR (SA$^2$)

The Spectrum Analyzer and Animator (SA$^2$), whose interface appears on Figure 3, automates the following procedure to generate a sequence of synthetic binary wind-line profiles to fit a sequence of observed wind-line profiles.

1. *Generation of single-star SEI profiles*: The stellar components of the binary are first "fitted" with individual single-star winds, which are assumed to be centered on their host stars. These steady-state, time-independent individual wind structures yield phase-independent single-star SEI profiles $F_1(\lambda)$ and $F_2(\lambda)$ for the primary and secondary star of the binary, respectively. Because a single-star SEI profile $F(\lambda)$ is the sum of the contributions for impact parameters impinging the photosphere (p < 1) and impact parameters not-impinging the photosphere (p ≥ 1), then for star j=1,2:

$$F_j(\lambda) \equiv F_j(\lambda; p \geq 0) = F_j(\lambda; p < 1) + F_j(\lambda; p \geq 1)$$

2. *Doppler-shift of SEI profiles*: For every IUE spectrum at orbital phase φ, this second step entails the Doppler-shift of the individual phase-independent SEI profiles using radial velocities $RV_1(\varphi)$ and $RV_2(\varphi)$ for the stellar components that follow from published radial velocity solutions. This process yields the individual phase-dependent profiles $F_{\{1,2\}}(\lambda;\varphi)$:

$$F_j(\lambda;\phi) = F_j(\lambda) \cdot (1 + RV_j(\phi)/c) = [F_j(\lambda; p < 1) + F_j(\lambda; p \geq 1)] \cdot (1 + RV_j(\phi)/c),$$

where c is the speed of light.

3. *Amalgamation of SEI profiles*: In this third step, these phase-dependent profiles are "combined" at each phase φ into a phase-dependent synthetic Binary-SEI, or BSEI profile F(λ;φ), taking into account (a) the contribution of each stellar photosphere into UV continuum flux and (b) the published intrinsic photospheric light ratio $L_1/L_2$ of the member stars in the UV which is 0.6/0.4 (Stickland, Koch, Pachoulakis and Pfeiffer, 1993). The weighing scheme adopted in the calculation of the synthetic BSEI profile F(λ;φ) assumes disparate winds for the components of the binary that are not seriously deformed from spherically symmetric shapes due to e.g., severe wind collisions.

$$F(\lambda;\phi) \equiv F(\lambda;\phi; p \geq 0) =$$
$$= [W_1 \cdot F_1(\lambda; p < 1) + L_1 \cdot F_1(\lambda; p \geq 1)] \cdot (1 + RV_1(\phi)/c) +$$
$$[W_2 \cdot F_2(\lambda; p < 1) + L_2 \cdot F_2(\lambda; p \geq 1)] \cdot (1 + RV_2(\phi)/c)$$

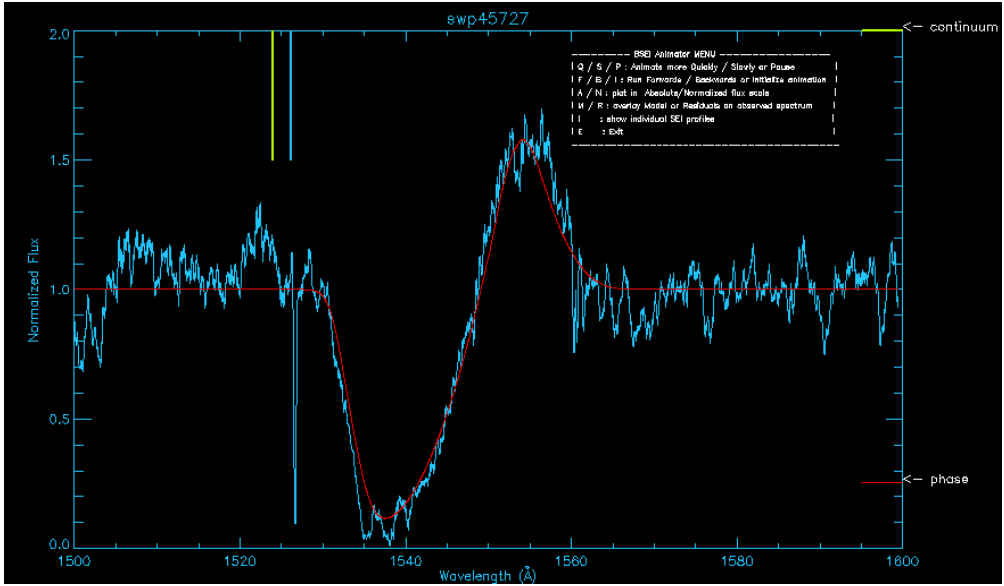

**Figure 3: The interface of the Spectrum Analyzer and Animator (SA²) for wind line profile modeling.**

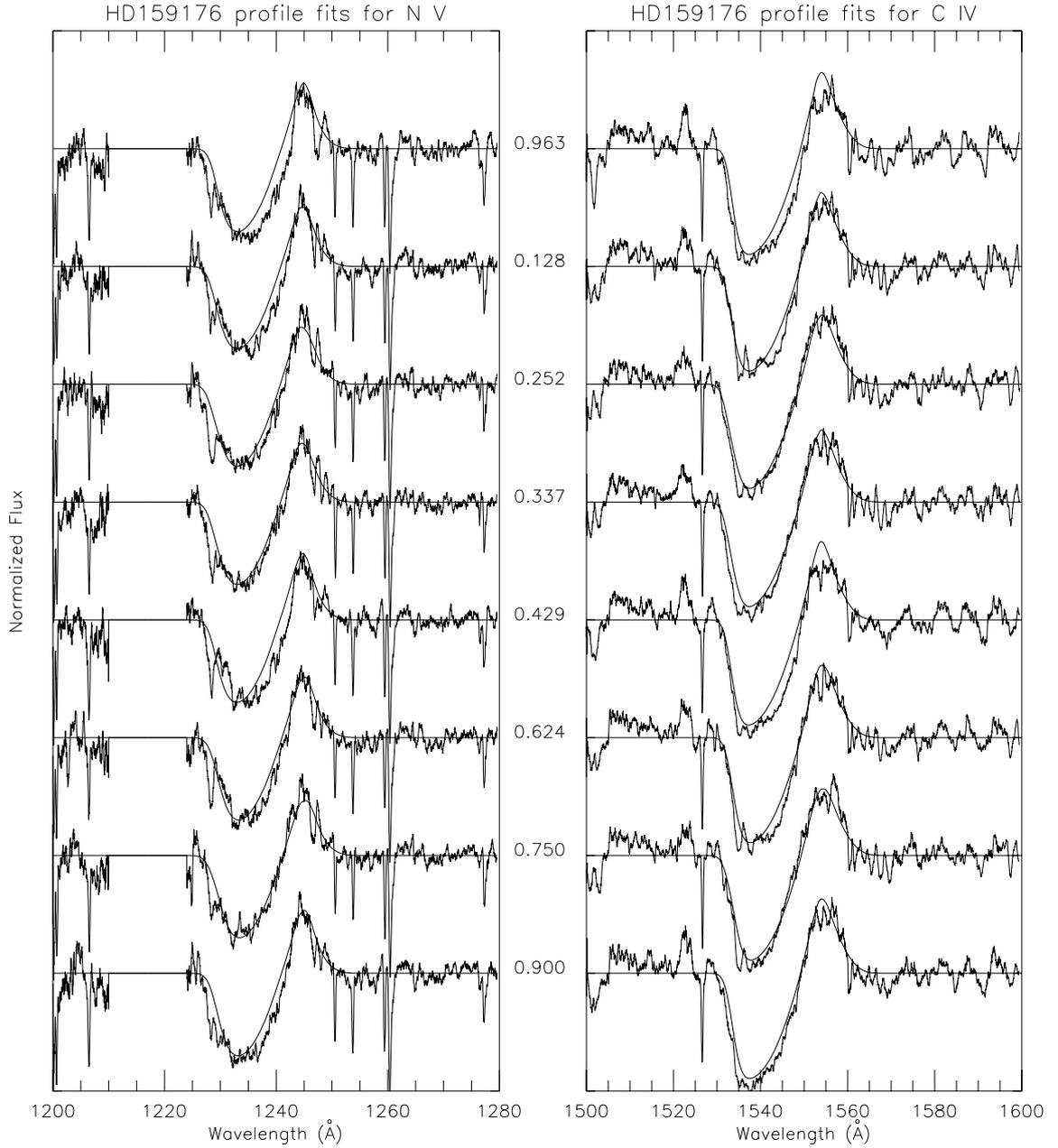

**Figure 4: Output of SA² showing the BSEI fits to the observed NV and CIV profiles.**

For phases outside photospheric eclipses, $[W_1, W_2] = [L_1, L_2]$, so that the Doppler-shifted SEI profiles (for both $p < 1$ and $p \geq 1$) are weighed according to the intrinsic photospheric light ratio. For phases during photospheric eclipses, $[W_1, W_2] = [LC-L_1, L_2]$ if the primary star is eclipsed and $[W_1, W_2] = [L_1, LC-L_2]$ if the secondary star is eclipsed, where LC is the normalized flux measure at that phase derived from the continuum light curves. It is to be noted that a weighing scheme for the $p \geq 1$ contributions cannot be devised because intrinsic "wind" weights are not known and the wind-affected light curve cannot be used as it contains total wind $p \geq 0$ contributions. Accordingly, the weights for the $p \geq 1$ contributions have been kept constant throughout the Keplerian cycle. This procedure is justified because (a) the wind models for these binaries lead to flow times through the line-forming regions of about 2-5% of the Keplerian period and (b) the exposure times of the raw spectral images are only about 30% of the flow time.

Using continuum and wind-affected light curves from Pachoulakis (1996), the observed wind-line profiles are normalized to a continuum flux level of 1.0, so that they can be compared to the synthetic binary wind-line profiles. Finally, truncated spectral intervals centered on the observed wind-line are played back on screen in phase order with their corresponding model line-profiles $F(\varphi)$ superimposed in order to assess the quality of the fit (see Figure 3). The sequence of steps just outlined and beginning with the calculation of new model SEI profiles is iterated until a steady-state composite wind is computed that satisfactorily represents the observed wind-line profiles for each ion.

Figure 4 displays the resultant composite-wind BSEI profiles for the dominant wind lines of HD159176. When perusing this figure, the reader should keep in mind that the phase-locked variabilities of the BSEI model profiles arise solely from the phase-dependent Doppler shifts and the phase-dependent weighting scheme adopted for their creation and represent a steady-state binary wind. The reader may also note that better fits have been attained near quadrature phases than near conjunction phases for both NV and CIV. In fact, as can be seen from Figure 4, these two phases represent the two extremes in quality of fit among all available spectra: the remaining spectra are of intermediate character which varies with phase in a more or less continuous manner between conjunctions and quadratures. The regularity and phase-locked pattern of the fitting quality attests to variability caused by the phase-dependent wind geometry of the binary as observed from Earth.

## 4. Discussion

A sequence of articles, of which the present is the first, reports on a method used to model, simulate and visualize the 3D structure of astrophysical wind volumes, which operates in a way similar to multi-directional medical tomography in that the spatial structure of an extended target can be reconstructed from a number of spectra obtained by scanning that target from several directions. It is precisely the natural experiments of Keplerian motion and the interactions arising from the presence of a stellar companion that provide the evidence to unlock the structure of each component wind most informatively. Depending on the orientation of the orbit, as the stars trace their Keplerian orbits, one star/wind may be occulted/attenuated by the other star/wind. Accordingly, each photospheric disk – and to a much lesser degree each wind – acts as an analyzing source of light to the front wind. The winds of HD159176 are sufficiently extended and optically thick to deform the wind lines of NV and of CIV into P-Cygni profiles, unmistakable diagnostics of out-flowing material.

The modeling methodology outlined in the previous sections uses a custom piece of software, the Spectrum Analyzer and Animator ($SA^2$) to obtain a physically consistent solution for the binary wind which represents a steady-state composite wind with (possibly) transient features such as wind inhomogeneities in the circumstellar wind volume. The $SA^2$ brings together a number of custom-made routines under a common interface and operates in a number of modes to allow the following overall functionalities: extract light curves at continuum and wind-affected bandpasses, normalize the observed spectra and produce synthetic binary wind line (BSEI) profiles which reproduce the observed wind line variabilities.